\documentstyle[12pt]{article} 
\newcommand{\boldC}{\bf C}

\newcommand{\CGC}{\mbox{$C^{\infty}_{c}(G,{\bf C})$}}

\newcommand{\DerA}{\mbox{Der${\cal A}$}}

\newcommand{\calE}{\mbox{${\cal E}$}}

\newcommand{\Cstar}{\mbox{$C^*$}}

\newcommand{\calA}{\mbox{${\cal A}$}}
\newcommand{\calB}{\mbox{${\cal B}$}}
\newcommand{\calH}{\mbox{${\cal H}$}}
\newcommand{\Aproj}{\mbox{${\cal A}_{proj}$}}
\newcommand{\calAG}{\mbox{$\calA_{\Gamma }$}}

\newcommand{\calAGamma}{\mbox{${\cal A}_{\Gamma }$}}

\begin{document}

\title{Einstein-Podolski-Rosen Experiment 
from Noncommutative Quantum Gravity}

\author{%
Michael Heller \\
Vatican Observatory, V-12000 Vatican City
State\thanks{Correspondence address: ul.  Powsta\'nc\'ow
Warszawy 13/94, 33-110 Tarn\'ow, Poland. E-mail:
mheller@wsd.tarnow.pl}
\and
Wies{\l}aw Sasin \\ Institute of Mathematics, Warsaw University
of Technology \\ Plac Politechniki 1, 00-661 Warsaw, Poland}

\maketitle

\begin{abstract}
It is shown that the Einstein--Podolski--Rosen type experiments
are the natural consequence of the groupoid approach to
noncommutative unification of general relativity and quantum
mechanics. The geometry of this model is determined by the
noncommutative algebra $\calA = \CGC $ of complex valued,
compactly supported, functions (with convolution as
multiplication) on the groupoid $G = E \times \Gamma $. In the
model considered in the present paper $E$ is the total space of
the frame bundle over space-time and $\Gamma $ is the Lorentz
group. The correlations of the EPR type should be regarded as
remnants of the totally non-local physics below the Planck
threshold which is modelled by a noncommutative geometry.
\end{abstract}
\par
\newpage

\section{Introduction}
One of the greatest challenges of contemporary physics is to
explain the non-local effects of quantum mechanics theoretically
predicted (in the form of a gedanken experiment) by Einstein,
Podolski and Rosen (EPR, for short) \cite{EPR} and
experimentally verified by Aspect et al.
\cite{Aspect1,Aspect2,Aspect3} (for a comprehensive review see
\cite{Redhead}). Although non-local effects of this type
logically follow from the postulates of quantum mechanics, it
seems strange and against our "realistic common sense" to accept
that two particles separated in space could be so strongly
correlated (provided they once interacted with each other) that
they seem to ``know'' about each other irrespectively of the
distance separating them. In spite of long lasting discussions,
so far no satisfactory explanation of this effect has been
offered. In the present paper we shall argue that effects of the
EPR type are remnants of the totally non-local physics of the
fundamental level (below the Planck threshold). We substantiate
our argument by explaining the EPR experiment in terms of a
quantum gravity model, based on a noncommutative geometry,
proposed by us in \cite{HSL} (see also
\cite{HS1}), although the explanation itself does not depend on
particulars of the model.
\par
The main physical idea underlying our model is that below the
Planck threshold (we shall speak also on the "fundamental
level") there is no space-time but only a kind of pregeometry
which is modeled by a suitable noncommutative space, and that
space-time emerges only in the transition process to the
classical gravity regime.  Accordingly, we start our
construction not from a space-time manifold $M$, but rather from
a groupoid $G = E \times \Gamma $ where $E$ is a certain
abstract space and $\Gamma $ a suitable group of "fundamental
symmetries". In the present paper, for the sake of simplicity,
we shall assume that $E$ is the total space of the frame bundle
over space-time $M$ and $\Gamma = SO(3,1)$.  We define, in terms
of this geometry, the noncommutative algebra $\calA = \CGC $ of
smooth, compactly supported, complex-valued functions on the
groupoid $G$ with the usual addition and convolution as
multiplication. We develop a noncommutative differential
geometry basing on this algebra, and define a noncommutative
version of Einstein's equation (in the operator form). The
algebra \calA \ can be completed to become a
\Cstar-algebra, and this subalgebra of
\calA \ which satisfies the generalized Einstein's equation is
called {\it Einstein \Cstar- algebra}, denoted by \calE \ (for
details see \cite{HSL}). And now quantization is performed in
the standard algebraic way.  Since the explanation of the EPR
type experiments depends on the noncommutative structure of the
groupoid $G$ rather than on details of our field equations and
the quantization procedure, we shall not review them here; the
reader interested in the particulars of our model should consult
the original paper \cite{HSL}.
\par
It can be shown that the subalgebra \Aproj \ (elements of \Aproj
\ are called {\em projectible\/}) of functions which are
constant on suitable equivalence classes of fibres
$\pi_E^{-1}(p)$, $\pi_E$ being the projection $G=E\times
\Gamma \rightarrow \Gamma $ and $p \in E$, is isomorphic to the
algebra $C^{\infty }(M)$ of smooth functions on $M$.
Consequently, by making the restriction of \calA \ to \Aproj \
we recover the ordinary space-time geometry and the standard
general relativity. In our model, to simplify calculations, we
have assumed that the noncommutative differential geometry is
determined by the submodule $V$ of the module \DerA \ of all
derivations of \calA , and that $V$ has the structure adapted to
the product structure of the groupoid $G=E\times \Gamma $, i.
e., $V = V_E \oplus V_{\Gamma }$, where $V_E$ and $V_{\Gamma }$
are "parts" parallel to $E$ and $\Gamma $, correspondingly. It
can be seen that in our model the geometry ``parallel'' to $E$
is responsible for generally relativistic effects, and that
``parallel'' to $\Gamma $ for quantum effects. In general,
"mixed terms" should appear, and then one would obtain stronger
interaction between general relativity and quantum physics. This
remains to be elaborated in the future.
\par
The crucial point is that the geometry as determined by the
noncommutative algebra \calA \ is non-local, i. e., there are no
maximal ideals in \calA \ which could determine points and their
neighborhoods in the corresponding space, and consequently
neither space points nor time instants can be defined in terms
of \calA . Physical states of a quantum gravitational system are
identified with states on the algebra \calA , i. e., with the
set of positive linear functionals (normed to unity) on \calA ,
and pure states in the mathematical sense are identified with
pure states in the physical sense. Let $a \in \cal A $ be a
quantum gravitational observable, i. e., a projectible and
Hermitian element of \calA \ ($a$ must be projectible to leave
traces in the macroscopic world), and $\varphi $ a state on
\calA. Then $\varphi (a)$ is the expectation value of the
observable $a$ when the system is in the state $\varphi $. The
fact that $a$ is an element of a "non-local" (noncommutative)
algebra \calA \ implies that when $a$ is projected to the
space-time $M$ it becomes a real-valued (since $a$ is Hermitian)
function on $M$, and the results of a measurement corresponding
to $a$ are values of this function.  Consequently, one should
expect correlations between various measurement results even if
they are performed at distant points of space-time $M$. We shall
see that this is indeed the case.
\par
The organization of our material is the following. In Section 2,
we consider the eigenvalue equation for quantum gravitational
observables. In Section 3, we show that correlations of the EPR
type between distant events in space-time are consequences of
non-local (noncommutative) physics of the quantum gravitational
regime, and in Section 4 we present details of the EPR
experiment in terms of the noncommutative approach. Section 5
contains concluding remarks.

\section{Measurement on Quantum Gravitational System}
Let $\varphi : \calA \rightarrow \boldC $ be a state on the
algebra \calA , i. e., $\varphi({\bf 1})=1$ and $\varphi (aa^*)
\geq 0$ for every $a \in \calA $. It can be easily seen that
$\varphi|\calA_{proj}: \calA_{proj} \rightarrow \boldC $ is a
state on the subalgebra $\calA_{proj}$.
\par
Let now $a \in \calA_{proj} $ be Hermitian, then there exists a
function $\bar{a} \in C^{\infty }(M)$ with $ \bar{a} \circ pr =
a$, where $pr: G \rightarrow M$ is the projection, and the state
$\bar{\varphi }: C^{\infty }(M) \rightarrow {\bf R}$ on the
algebra $C^{\infty }(M)$, such that  $\varphi (a) = \bar{\varphi
}(\bar{a})$. Since the algebras $\calA_{proj}$ and $C^{\infty
}(M)$ are isomorphic, the spaces of states of these algebras are
isomorphic as well.
\par
To make a contact with the standard formulation of quantum
mechanics we represent the noncommutative algebra \calA \ in a
Hilbert space by defining, for each $p\in E$, the representation
$$
\pi_p: \calA \rightarrow \calB (\calH ),
$$
where $\calB (\calH )$ is the algebra of operators on the
Hilbert space $\calH = L^2(G_p)$ of square integrable functions
on the fibre $G_p = \pi_E^{-1}(p)$, $\pi_E: G \rightarrow E$
being the natural projection, with the help of the formula
$$
\pi_p(a)\psi = \pi_p(a)*\psi
$$
or more explicitly
$$ 
(\pi_p(a)\psi)(\gamma ) = \int_{G_p}a(\gamma_1)\psi(\gamma_1^{-
1}\gamma),
$$
$a \in \calA,\, \gamma = \gamma_1 \circ \gamma_2,\; \gamma ,
\gamma_1, \gamma_2 \in G_p$, $\psi \in L^2(G_p)$, and the
integration is with respect to the Haar measure. This
representation is called the {\em Connes representation\/} (see
\cite[p.102]{Connes}, \cite{HSL}).
\par
Now, let us suppose that $a$ is an observable, i. e., $a \in
\calA_{proj} $, and we perform a measurement of the quantity
corresponding to this observable. The eigenvalue equation for
$a$ is
\begin{equation} 
\int_{G_P}a(\gamma_1)\psi(\gamma_1^{-1}\gamma ) = r_p\psi
(\gamma) \label{eq1} \end{equation}
where the eigenvalue $r_q$ is the expected result of the
measurement when the system is in the state $\psi $. Here we
must additionally assume that the "wave function" $\psi $ is
constant on fibres of $G$ to guarantee for equation (\ref{eq1})
to have its usual meaning in the non-quantum gravity limit. If
this is the case, equation (\ref{eq1}) can be written in the
form
$$
\psi (\gamma_1^{-1}\gamma ) \int_{g_p}a(\gamma_1) = r_p\psi
(\gamma )
$$
and consequently
$$
r_p = \int_{G_p}a(\gamma_1).
$$
Let us notice that the measurement result is a measure in the
mathematical sense. 
\par
It is obvious that if we define the "total phase space" of our
quantum gravitational system
$$
L^2(G) := \bigoplus_{p\in E}L^2(G_p)
$$
and the operator
$$
\pi(a) := (\pi_p(a))_{p\in E}
$$
acting on $L^2(G)$ then the eigenvalue equation becomes
$$
\pi (a) \psi = r\psi 
$$
where $r: M \rightarrow {\bf R}$ is a function on space-time $M$
given by
\begin{equation}
r(x) = \int_{G_p}a(\gamma_1)
\label{eq2} \end{equation}
where $x$ is a point in $M$ to which the frame $p$ is attached.
Let us notice that the function $r$ is equal to the function
$\bar{a}: M \rightarrow {\bf R}$ (see the beginning of the
present Section). Let us now consider a composed quantum system
the state of which is described by the single vector in the
Hilbert space, and let us perform a measurement on its parts
when they are at a great distance from each other. Formula
(\ref{eq2}) asserts that in such a case the results of the
measurement are not independent but are the values of the same
function defined on space-time.  This can be regarded as a
"shadow" of a non-local character of the observable $a$
projected down to space-time $M$.
\par

\section{EPR Non-Locality}
So far we were mainly concerned with what happens when we
project the algebra \calA \ onto the ``horizontal component''
$E$ of the groupoid $G$. This, of course, gives us the
transition to the classical space-time geometry (general
relativity). In the present Section, we shall be interested in
projecting \calA \ onto the ``vertical component'' $\Gamma $ of
$G$. This gives us quantum effects of our model.
\par
Let us consider functions {\em projectible\/} to $\Gamma $. We
define 
$$
\calAGamma := \{f \circ pr_{\Gamma }: f\in C^{\infty }_c(\Gamma,
\boldC) \} \subset \calA.
$$
The reasoning similar to that in the beginning of the present
section shows that if $s \in \calAGamma $ and $\psi : \calAGamma
\rightarrow \boldC $ is a state on \calAGamma \ then $\psi (s) =
\underline{\psi }(\underline{s})$, where $s
= \underline{s} \circ pr_{\Gamma }, \; pr_{\Gamma }: G
\rightarrow \Gamma $ is the projection, and $\underline{\psi }:
C^{\infty }_c(\Gamma, \boldC ) \rightarrow \boldC $ is a state
on $C^{\infty }_c(\Gamma, \boldC )$.
\par
Let now $\Phi $ be a state on $C^{\infty }_c(\Gamma, \boldC )$.
We say that the state $\varphi: \calA \rightarrow \boldC $ is
{\em $\Gamma $-invariant associated to $\Phi $\/} on \calA \ if
\[
\varphi (s) = \left\{
\begin{array}{cr}
\Phi(\underline{s}), & \mbox{if $s \in \calAGamma $},\\
0, &   \mbox{if $s \not\in \calAGamma$ }.
\end{array}
\right.
\]
Since all fibres of $G_p, \, p \in E$, of $G$ are isomorphic,
the number $\varphi (s) = \Phi (\underline{s})$, for $s \in
\calAGamma $, is the same in each fibre $G_p$. If additionally
$s$ is a Hermitian element of \calA , and if a measurement
performed at a certain point of space-time $M$ gives the number
$\varphi (s)$ as its result, then this result is immediately
``known'' at all other fibres $G_p$, $p\in E$, of $G$, and
consequently at all other points of space-time $x = \pi_M(p) \in
M$, where $\pi_M: E \rightarrow M$ is the canonical projection.
\par
This can be transparently seen if we consider the problem in the
Hilbert space by using the Connes representation of the algebra
\calA . Let $a \in \calAG $, and let us consider the
following Connes representations
\begin{equation}
\pi_p(a)(\xi_p) = a_p * \xi_p,
\label{Connesa} \end{equation}
and
\begin{equation}
\pi_q(a)(\xi_q) = a_q * \xi_q,
\label{Connesb} \end{equation}
where $\xi_p \in L^2(G_p), \, \xi_q \in L^2(G_q), \, p,q \in E,
\, p \neq q$. Since $G_p$ and $G_q$ are isomorphic, we can
choose $\xi_p$ and $\xi_q$ to be isomorphic with each other,
which implies that $\pi_p(a)$ and $\pi_q(a)$ are isomorphic as
well. We have the following important
\par
{\sl Lemma.} If $a \in \calAG $ then its image under the Connes
representation $\pi_p$ does not depend of the choice of $p \in
E$ (up to isomorphism).
\par
Since $p \in E$ projects down to the space-time point $\pi_M(p)
\in M$, $\pi_M: E \rightarrow M$, the above result should be
interpreted as stating that all points of $M$ ``know'' what
happens in the fiber $G_g, \, g \in \Gamma $. This, together
with the fact that vectors $\xi_p$, upon which the observable
$\pi_p(a)$, $a \in \calA_{\Gamma }$ acts, also do not depend of
$p$, in principle, explains the EPR type experiments. However,
let us go deeper into details.
\par

\section{EPR Experiment in Terms of Noncommutative Geometry}
In this section we consider a group $\Gamma $ such that
$\Gamma_0= SU(2)$ is its compact subgroup. We look for an element
$s\in
\calAG$ such that
$$
\pi_p(s): L^2(\Gamma_0) \rightarrow L^2(\Gamma_0).
$$
Of course, $\boldC^2 \subset L^2(\Gamma_0)$. We define two
linearly independent functions on the group $\Gamma_0$, for
instance the constant function
$$
{\bf 1}: \Gamma_0 \rightarrow \boldC,
$$
and
$$
{\rm det}: \Gamma_0 \rightarrow \boldC ,
$$
which span the linear space $\boldC^2$, i. e., $\boldC^2 =
\langle {\bf 1}, {\rm det}\rangle_{\boldC }$. Let $\hat{S}_z =
\pi_p(s)|_{\boldC^2}$ be the usual z-component spin operator.
We have 
$$
\pi_p(s)\psi = \hat{S}_z\psi ,
$$
for $\psi \in \boldC^2$ or, by using the Connes representation
and the fact that 
$\hat{S}_z\psi = \pm \frac{\hbar }{2}\psi $,
$$
\int_{\Gamma_0}s_p(\gamma_1)\psi (\gamma_1^{-1}\gamma) = \pm
\frac{\hbar}{2}\psi .
$$
Since $s_p = $const, one obtains
$$
\int_{\Gamma_0 }\psi (\gamma_1^{-1} \gamma ) \sim
\psi (\gamma ).
$$
One of the solutions of this equation is $\psi = {\bf
1}_{\Gamma_0}$. Therefore
$$
\frac{\hbar }{2} = \pm \int_{\Gamma_0}s_p(\gamma_1).
$$
Hence
$$
(s_p)_1 = +\frac{\hbar }{2} \frac{1}{{\rm vol}\Gamma_0},
$$ $$
(s_p)_2 = -\frac{\hbar }{2} \frac{1}{{\rm vol}\Gamma_0},
$$
and consequently
$$
\pi_p((s_p)_1)\psi = + \frac{\hbar }{2}\psi  \;\; {\rm for}\,
\psi\in \boldC^+, 
$$ $$
\pi_p((s_p)_2)\psi = - \frac{\hbar }{2}\psi  \;\; {\rm for}\,
\psi\in \boldC^-, 
$$
where $\boldC^+ := \boldC \times \{0\}$, and $\boldC^- :=
\{0\} \times \boldC$. To summarize these results we can define
$$
\hat{S}_z \psi = \pi_p(s_1,s_2)\psi := 
\left\{
\begin{array}{cr}
(s_{1})_p*\psi & \mbox{{\rm if} $\psi \in \boldC^+$,} \\
(s_{2})_p*\psi  & \mbox{{\rm if} $\psi \in \boldC^-$.}
\end{array}
\right.
$$
\par
Now, the analysis of the ``EPR paradox'' proceeds in the same
way as in the standard textbooks on quantum mechanics (see for
instance \cite[pp. 179-181]{Isham}. An observer $A$, situated at
$\pi_M(p) = x_A \in M$, measures the z-spin component of the one
of the electrons\footnote{Let us notice that when $A$ measures
the spin of the electron, he simultaneously determines the
position of the electron (at least roughly), i. e., the position
$x_A$ at which he himself is situated (spin and position
operators commute).}, i. e., he applies the operator $\hat{S}_z
\otimes {\bf 1}|_{\boldC^2}$ to the vector $\xi =
\frac{1}{\sqrt{2}}(\psi \otimes
\varphi - \varphi \otimes \psi )$ where $\psi \in \boldC^+$ and
$\varphi \in \boldC^-$. Let us suppose that the result of the
measurement is $\frac{\hbar }{2}$. This means that the state
vector $\xi = \frac{1}{\sqrt{2}}(\psi \otimes \varphi -
\varphi \otimes \psi ) \in {\boldC}^2 \otimes {\boldC}^2 \subset
L^2(G_r) \otimes L^2(G_r)$, $r \in E$, has collapsed to $\xi_0
=\frac{1}{\sqrt{2}}(\psi \otimes \varphi )$, and that
immediately after the measurement the system is in the state
$\xi_0$ which is the same (up to isomorphism) for all fibres
$G_r$ whatever $r \in E$, and consequently it does not depend of
the point in space-time to which $r$ is attached (see formulae
(\ref{Connesa}) and (\ref{Connesb}) which are obviously valid
also for tensor products). In particular, the vector $\xi_0$ is
the same for the fibres $G_p$ and $G_q$ where $p$ is such that
$\pi_M(p) = x_A$ and $q$ is such that $\pi_M(q) = x_B \; (x_A
\neq x_B)$. It is now obvious that if an observer $B$, situated
a $x_B$  measures the z-spin component of the second electron,
i. e., if he applies the operator $ {\bf 1}|_{\boldC^2} \otimes
\hat{S}_z$ to the vector $\xi_0$, he will obtain the value
$-\frac{\hbar }{2}$ as the result of his measurement.
\par

\section{Concluding Remarks}
To conclude our analysis it seems suitable to make the following
remarks.
\par
It should be emphasized that our scheme for noncommutative
quantum gravity does not ``explain'' quantum mechanical
postulates. In the very construction of our scheme it has been
assumed that the known postulates which, in the standard
formulation of quantum mechanics are valid for the algebra of
observables, can be extended to a more general noncommutative
algebra. However, the very fact that these postulates are valid
in the conceptual framework of noncommutative geometry gives
them a new flavour. For instance, since in the noncommutative
regime there is no time in the usual sense, the sharp
distinction between the continuous unitary evolution and the
non-continuous process of measurement (``collapse of the wave
function'') disappears. This distinction becomes manifest only
when time emerges (see \cite{Emergence}) in changing from the
noncommutative regime to the usual space-time geometry.
\par
What our approach does explain is the fact that some quantum
effects are strongly correlated even if they occur at great
distances from each other. These effects are ``projections''
from the fundamental level at which all concepts have purely
global meanings.
\par
This explanation does not depend on ``details'' of our model,
such as some particulars of the construction of noncommutative
differential geometry, the concrete form of generalized
Einstein's equation, or the dynamical equation for quantum
gravity. However, it does depend on (or even more, it is deeply
rooted in) the noncommutative character of the algebra $\calA =
\CGC $ and the product structure of the groupoid $G = E
\times \Gamma $.
\par

\end{document}